\documentclass[aps,prl,amsmath,amssymb,reprint,superscriptaddress]{revtex4-2}
\usepackage{graphicx}
\usepackage{bm}
\usepackage{threeparttable}

\begin{document}

\title{Electro-optically Modulated Nonlinear Metasurfaces}

\author{Zhengqing He $^{\ddag}$}
\affiliation{The Key Laboratory of Weak-Light Nonlinear Photonics, Ministry of Education, School of Physics and TEDA Applied Physics Institute, Nankai University, Tianjin 300071, People’s Republic of China}

\author{Lun Qu $^{\ddag}$}
\affiliation{The Key Laboratory of Weak-Light Nonlinear Photonics, Ministry of Education, School of Physics and TEDA Applied Physics Institute, Nankai University, Tianjin 300071, People’s Republic of China}

\author{Wei Wu}
\affiliation{The Key Laboratory of Weak-Light Nonlinear Photonics, Ministry of Education, School of Physics and TEDA Applied Physics Institute, Nankai University, Tianjin 300071, People’s Republic of China}

\author{Jikun Liu}
\affiliation{The Key Laboratory of Weak-Light Nonlinear Photonics, Ministry of Education, School of Physics and TEDA Applied Physics Institute, Nankai University, Tianjin 300071, People’s Republic of China}

\author{Jingfei You}
\affiliation{The Key Laboratory of Weak-Light Nonlinear Photonics, Ministry of Education, School of Physics and TEDA Applied Physics Institute, Nankai University, Tianjin 300071, People’s Republic of China}

\author{Weiye Liu}
\affiliation{The Key Laboratory of Weak-Light Nonlinear Photonics, Ministry of Education, School of Physics and TEDA Applied Physics Institute, Nankai University, Tianjin 300071, People’s Republic of China}

\author{Lu Bai}
\affiliation{The Key Laboratory of Weak-Light Nonlinear Photonics, Ministry of Education, School of Physics and TEDA Applied Physics Institute, Nankai University, Tianjin 300071, People’s Republic of China}

\author{Chunyan Jin}
\affiliation{The Key Laboratory of Weak-Light Nonlinear Photonics, Ministry of Education, School of Physics and TEDA Applied Physics Institute, Nankai University, Tianjin 300071, People’s Republic of China}

\author{Chenxiong Wang}
\affiliation{The Key Laboratory of Weak-Light Nonlinear Photonics, Ministry of Education, School of Physics and TEDA Applied Physics Institute, Nankai University, Tianjin 300071, People’s Republic of China}

\author{Zhidong Gu}
\affiliation{The Key Laboratory of Weak-Light Nonlinear Photonics, Ministry of Education, School of Physics and TEDA Applied Physics Institute, Nankai University, Tianjin 300071, People’s Republic of China}

\author{Wei Cai}
\affiliation{The Key Laboratory of Weak-Light Nonlinear Photonics, Ministry of Education, School of Physics and TEDA Applied Physics Institute, Nankai University, Tianjin 300071, People’s Republic of China}

\author{Mengxin Ren}
\email{ren\_mengxin@nankai.edu.cn}
\affiliation{The Key Laboratory of Weak-Light Nonlinear Photonics, Ministry of Education, School of Physics and TEDA Applied Physics Institute, Nankai University, Tianjin 300071, People’s Republic of China}
\affiliation{Collaborative Innovation Center of Extreme Optics, Shanxi University, Taiyuan, Shanxi 030006, People’s Republic of China}

\author{Jingjun Xu}
\email{jjxu@nankai.edu.cn}
\affiliation{The Key Laboratory of Weak-Light Nonlinear Photonics, Ministry of Education, School of Physics and TEDA Applied Physics Institute, Nankai University, Tianjin 300071, People’s Republic of China}

\begin{abstract}
Tunable nonlinearity facilitates the creation of reconfigurable nonlinear metasurfaces, enabling innovative applications in signal processing, light switching, and sensing. This paper presents a novel approach to electrically modulate SHG from a lithium niobate (LN) metasurface, exploiting the electro-optical (EO) effect. By fabricating a nanohole array metasurface on a thin LN film and applying an electric field, we demonstrate the alteration of the material's refractive index, resulting in resonance shifts and modulation of SHG intensity at specific wavelengths. Our findings provide valuable insights for the development of electrically tunable nonlinear light sources, quantum optics, dynamic nonlinear holography, and nonlinear information processing.
\end{abstract}

\maketitle
\section{Introduction}
Metasurfaces, composed of meticulously engineered subwavelength-scale structures, stand at the forefront of optical technology, offering significant potential to redefine the landscape of photonic device miniaturization and functionality \cite{yu2011light,ni2012broadband,sun2012gradient,kildishev2013planar,neshev2018optical, guo2022classical}. This advancement is further accentuated by the infusion of nonlinear functionalities into metasurfaces, broadening their scope beyond the traditional linear optics \cite{kauranen2012nonlinear,lapine2014colloquium,li2017nonlinear}. Nonlinear metasurfaces, purposely designed to generate and control nonlinear phenomena such as second-harmonic generation (SHG), has catalyzed a plethora of groundbreaking applications encompassing nonlinear holography, optical encryption, and nonlinear beam shaping \cite{gao2018nonlinear,tang2019nonlinear,ye2016spin,mao2022nonlinear,keren2016nonlinear,wei2019efficient, hu2020nonlinear,keren2018shaping}. These significant strides in nonlinear metasurface research have propelled nonlinear flat optics as a pragmatic alternative to conventional nonlinear optics. Past decades have witnessed extensive endeavors to enhance SHG from metasurfaces. These efforts encompass techniques such as light field confinement through the excitation of plasmon resonances in metallic structures or Mie resonances in dielectric inclusions, aimed at amplifying SHG efficiency \cite{klein2006second,lee2014giant,o2015predicting,celebrano2015mode,liu2016resonantly,liu2019high, koshelev2020subwavelength,anthur2020continuous,timpu2019lithium,fang2020second,fedotova2020second, carletti2021steering,ma2021nonlinear,yuan2021strongly,qu2022giant,huang2022resonant,zhang2022quasi,zhao2023efficient,qu2023bright}. 

In addition to the pursuit of enhanced nonlinearity, achieving tunable nonlinearity is of paramount importance \cite{abdelraouf2022recent}. This capability enables the creation of reconfigurable nonlinear metasurfaces, facilitating novel applications in signal processing, light switching, and sensing. Researchers have diligently explored diverse methods to dynamically control SHG from metasurfaces using external stimuli. Notable approaches include electric-field-induced SHG \cite{cai2011electrically,kang2014electrifying,ding2014enhancement,chen2019gigantic}, and current-induced SHG \cite{takasan2021current} in plasmonic metasurfaces. Additionally, Stark tuning of intersubband transitions in quantum-well polaritonic metasurfaces has emerged as a successful method for controlling SHG in the infrared region \cite{yu2022electrically,chen2023electrically}. Furthermore, the integration of hybridized metasurfaces with liquid crystals has proven effective for tuning SHG performance, achieved through the manipulation of liquid crystal molecule orientation via external electric fields \cite{sharma2023electrically}.

In this paper, we demonstrate a new approach based on electro-optical (EO) effect in the metasurface to modulate its SHG. We identify lithium niobate (LN) as a particularly suitable material, owing to its large second order nonlinear susceptibilities showing synergistic combination of exceptional SHG and EO properties. We fabricated a nanohole array metasurface on a thin LN film, which exhibits distinct resonances in their transmission spectrum. By applying a specific electric field, the refractive index of the material is altered, causing a shift in the resonances. Consequently, this resonance shift impacts the intensity of the SHG signal at a specific wavelength, enabling the utilization of the electro-optical effect for SHG signal modulation. Our findings serve as valuable guidelines for innovative applications such as electrically tunable nonlinear light sources, quantum optics, dynamic nonlinear holography, and nonlinear information processing.

\section{Results}

\begin{figure}[htbp]
   \centering
  \includegraphics[width=90mm]{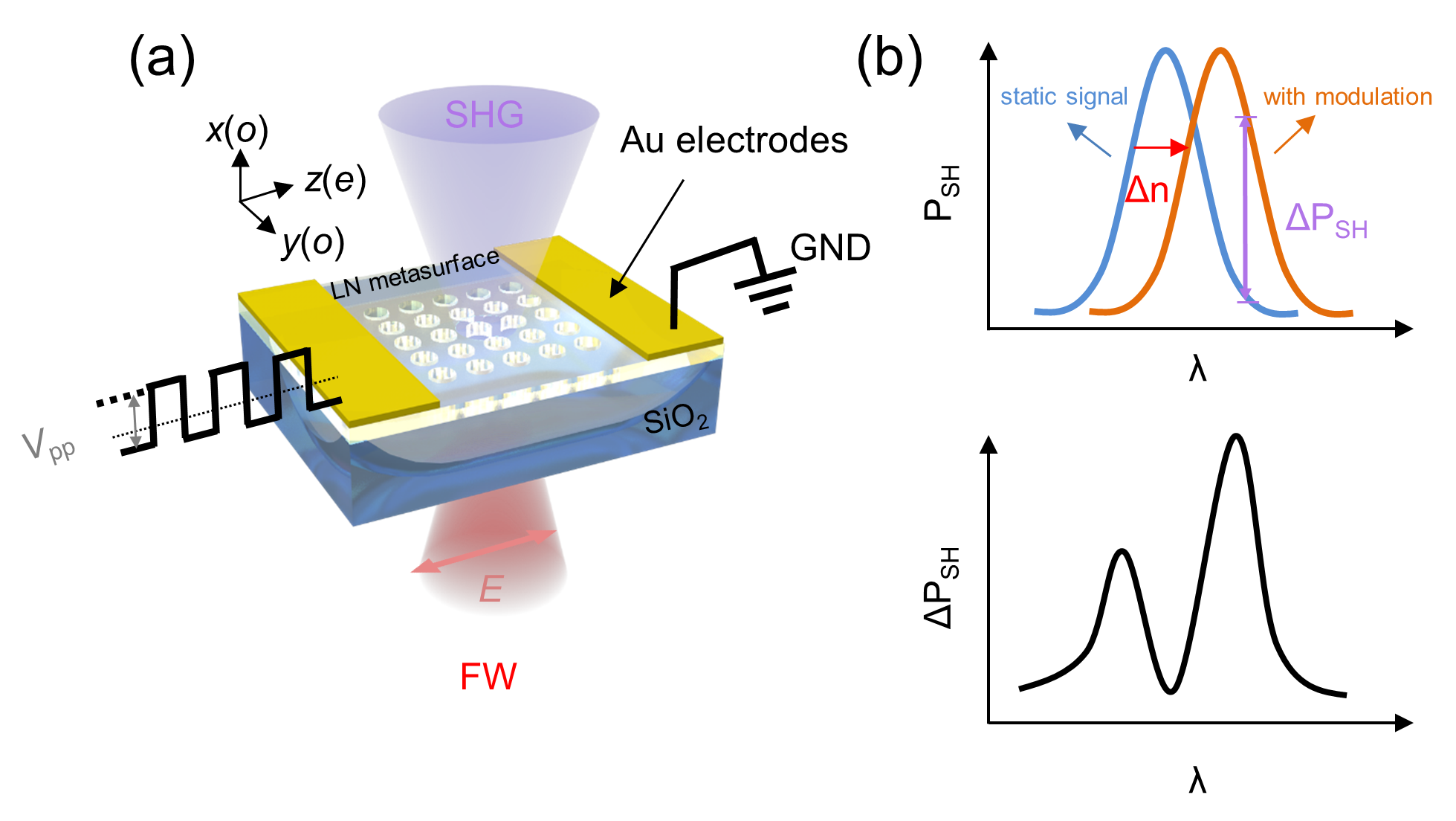}
  \caption{\textbf{ Concept of electrically modulated SHG from LN metasurface.} (a) The metasurface consists of a square lattice of air holes milled in an $x$-cut LN film. Experimental coordinates are chosen to overlap with the LN principal crystallographic axes. The $z$-direction corresponds to the optic axis. Two Au electrodes are positioned on the opposite side of metasurface array. A square alternating voltage wave with a peak-to-peak value of $V_{\text{pp}}$ is delivered into the electrodes. The incident FW light is $z$-polarized and normal to the metasurface, while the collected SHG signal is captured in the transmission direction. (b) Schematic diagram illustrating the principle of electrically modulated SHG. The wavelength dependence curve of the intensity of SHG ($P_{\text{SH}}$) before and after modulation is depicted above. The static signal (blue line) applies a voltage that causes a change in the refractive index of the sample, resulting in a shift in the wavelength domain (red line). The difference between the intensity of SHG before and after modulation at a given wavelength is the modulation amplitude ($\Delta$$P_{\text{SH}}$). The wavelength dependence curve of $\Delta$$P_{\text{SH}}$ is shown schematically in the figure below.
  }
  \label{fig1}
\end{figure}

The concept of electrically modulated SHG from a LN metasurface is illustrated in Figure 1a. The metasurface is composed of a square lattice of circular air holes etched into a suspended LN film, positioned within a gold electrode gap. SHG from the metasurface can be efficiently boosted by resonance, and wavelength of the maximum SHG efficiency coincides with the resonance peak \cite{qu2022giant}. To facilitate electrical modulation of the SHG, a square alternating voltage wave with a peak-to-peak value of $V_{\text{pp}}$ was fed into the electrodes. To capitalize the LN's largest second-order nonlinear susceptibilities, namely $d_{33}$ for SHG and $\gamma_{33}$ for EO modulation, the LN film is $x$-cut, and its optic axis (labeled by $z$-axis in Figure 1a) lies within the metasurface plane. Additionally, both the FW polarization and the modulating voltage were aligned along the $z$-axis. Under the influence of the external electric voltage, the refractive index of the LN undergoes alterations, resulting in a spectral shift of resonance. Due to the negative value of $\gamma_{33}$, the refractive index of the LN decreases under positive modulating voltage, causing a blue-shift in resonance, and vice versa (as depicted in Figure 1b). This resonance shift also leads to a shift in the SHG conversion efficiency spectrum. Consequently, the SHG magnitude at a specific wavelength experiences modulation. 
\begin{figure}[htbp]
  \centering
  \includegraphics[width=85mm]{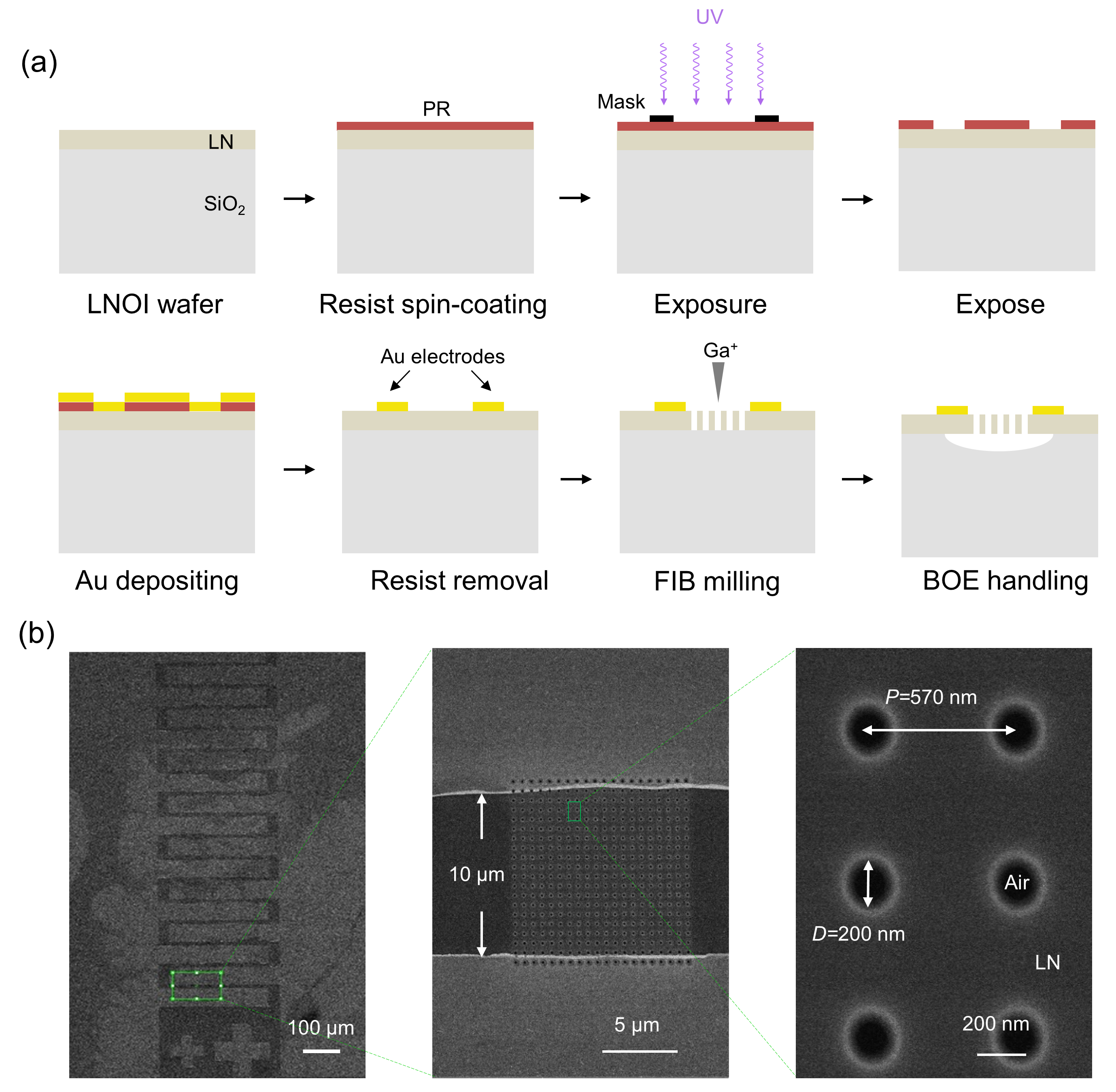}
  \caption{\textbf{Scanning electron microscopy (SEM) images of the LN metasurface.} (a) Overview of the key fabrication steps involved in creating the metasurface between Au electrodes. PR: photoresist, FIB: focused ion beaming, BOE: buffered oxide etching, IDT: interdigital electrodes. (b) Left: top-view SEM image depicting the Au interdigital electrodes (IDT) and the LN film within their slits. Middle: SEM image capturing the entire footprint of the metasurface, with an array size of 10$\times$10 $\mu$m$^2$. Right: zoomed-in SEM image revealing the detailed structure of the array, characterized by a period $P$=570~nm and air hole diameter $D$=200~nm. 
  }
  \label{fig2}
\end{figure}

Figure 2 depicts the fabrication process of the device. We started with an LNOI wafer, which consists of a 190~nm-thick crystalline LN film on a 300~$\mu$m-thick silicon dioxide (SiO$_2$) substrate. We employed a standard ultraviolet (UV) lithography technique to pattern an interdigital gold (Au) electrode array with 10~$\mu$m gaps, as depicted in the scanning electron microscope (SEM) image in Figure 2b. The metasurface was fabricated within the electrode gap using focused ion beam [FIB, gallium ions (Ga$^+$), 30 kV, 24 pA] milling. The overall footprint of the metasurface array was 10$\times$10 $\mu$m$^2$, featuring a lattice constant of 570~nm and an air hole diameter of 200 nm, as revealed in the SEM images in Figure 2b. Subsequently, the sample underwent immersion in a buffered oxide etching solution (BOE, NH$_4$F$\colon$HF=20$\colon$1) to selectively remove a micrometer-thin SiO$_2$ layer beneath the metasurface array. This process suspended the metasurface, offering advantages of increased refractive index contrast in the vertical direction and better light field confinement within the metasurface layer. 
\begin{figure}[htbp]
  \centering
  \includegraphics[width=85mm]{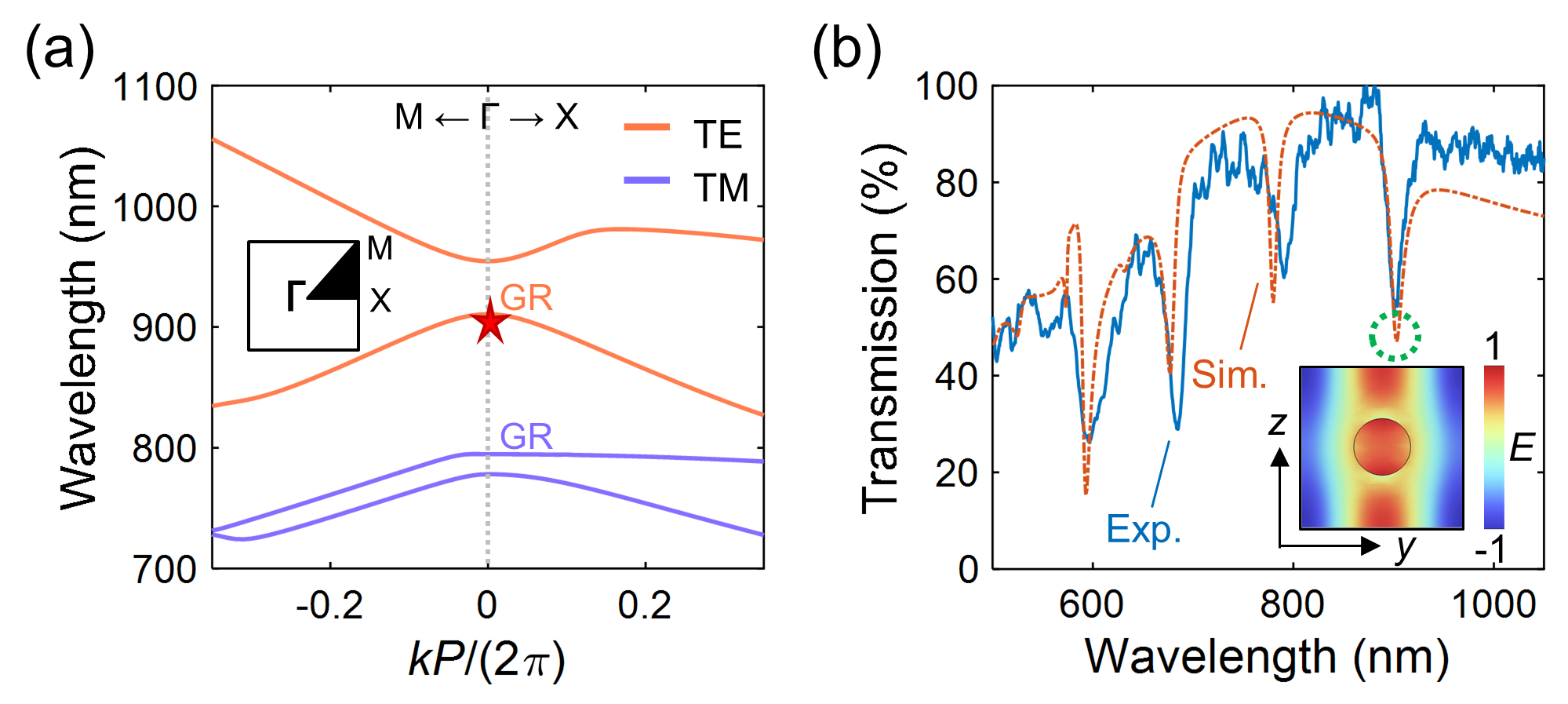}
  \caption{\textbf{Linear optical properties of the membrane metasurface.}  (a) Calculated band diagram of the metasurface. The symmetric points in an irreducible Brillouin zone are defined in the inset. The $\Gamma$-point at the center of the first Brillouin zone can be excited by normal incidence. The transverse magnetic (TM) and transverse electric (TE) modes are presented by purple and orange curves, respectively. (b) The experimental transmission spectrum under normally incident $z$-polarized light is shown by the blue curve, while the simulated transmission spectrum is depicted by the orange dashed line. The electric field map at a resonance wavelength of 902~nm (highlight by a green circle in the spectrum) is presented in the inset.
  }
  \label{fig3}
  \end{figure}

The metasurface displays discrete translational symmetry, a structural configuration linked with guided resonance modes \cite{fan2002analysis}. These modes efficiently confine electromagnetic energy within the metasurface layer, resulting in pronounced resonances in both the transverse magnetic (TM) and transverse electric (TE) bands, as shown by Figure 3a. The electric field of TE mode is primarily characterized by the $z$-component, facilitating effective interaction with the nonlinear coefficient $d_{33}$ of the LN. Therefore, our study focused on the TE mode. We employed the finite element method (COMSOL Multiphysics software) to simulate the transmission spectrum. The refractive index of LN used in the simulation was determined through ellipsometry (Accurion EP4) \cite{liu2021machine}. To account for optical losses caused by Ga ion implantation during the FIB fabrication, we introduced an extra imaginary part $\kappa$ of 0.015 to the refractive index of LN. The simulated spectrum, depicted by an orange dashed line in Figure 3b, reveal a prominent Fano-shaped resonance around 900~nm. This results in a significant localization of the electromagnetic field within the LN metasurface layer, as illustrated by the  electric field map in the inset. Experimental transmission spectrum was obtained using a commercial microscopic spectrometer (IdeaOptics Co., Ltd.), which is shown by blue line in Figure 3. The experimental result agrees reasonably with the simulation.
\begin{figure}[htbp]
  \centering
  \includegraphics[width=90mm]{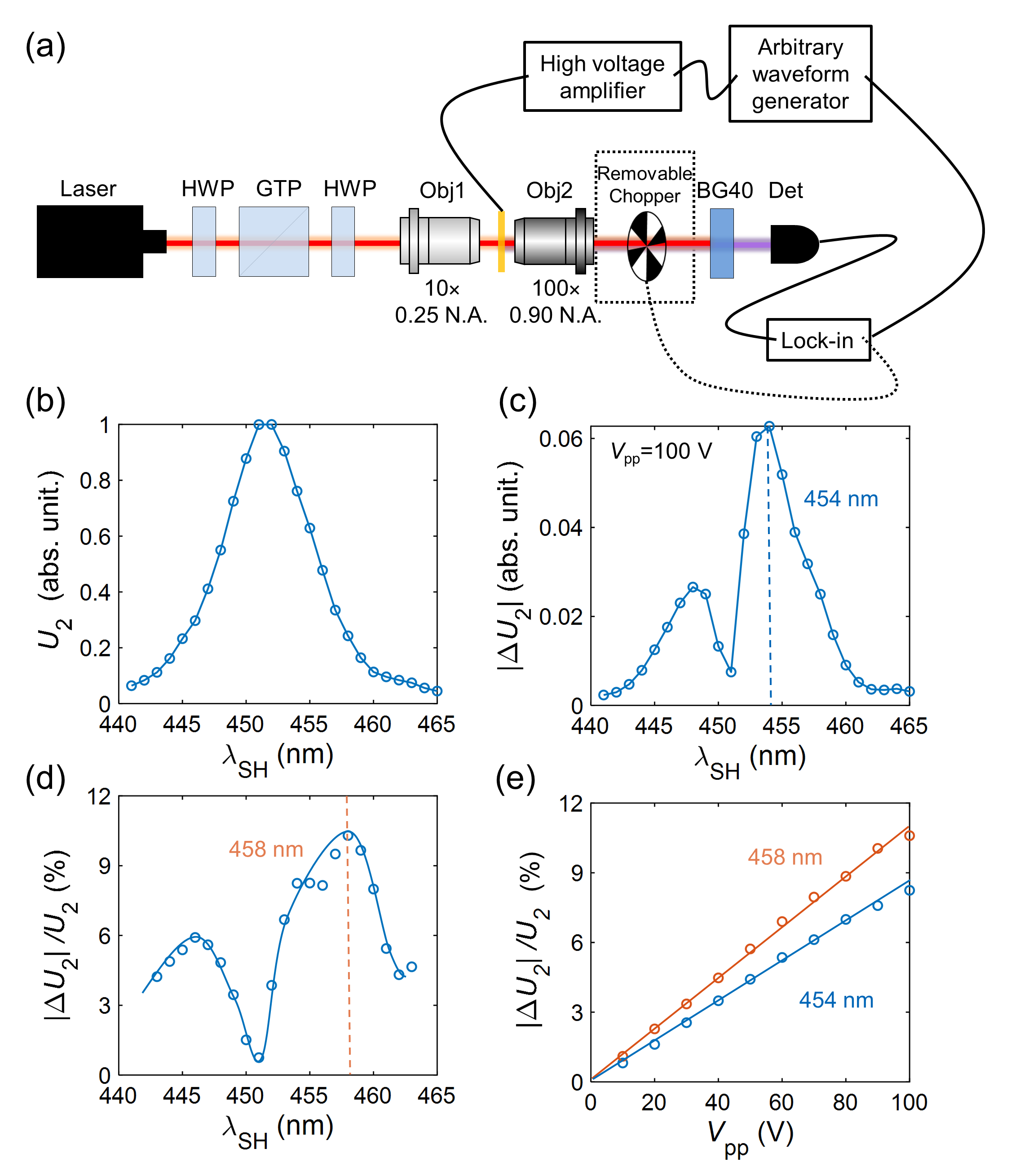}
  \caption{\textbf{Electrically modulated SHG from metasurface.} (a) Schematic representation of the experimental setup. Components include HWP (half-wave plate), GTP (Grant-Taylor prism), Obj (objective), BG40 (band-pass filter), and Det (balanced detector). (b) Experimentally measured wavelength dependent SHG strength, characterized by the detector output voltage $U_2$ in an arbitrary unit. (c) Electrically induced SHG strength alteration ($\vert$$\Delta $$U_2$$\vert$) under an external driving voltage with $V_{pp}$ of 100~V and frequency $f$ of 50~kHz. Circles represent the experimental data, with the line serving as an eye-guide. The maximum value is marked by the blue dash line. (d) Modulation depth, defined by $\vert$$\Delta$$U_2$$\vert$/$U_2$, as a function of pump wavelength. (e) Dependences of $\vert$$\Delta$$U_2$$\vert$/$U_2$ at 916~nm and 908~nm on $V_{pp}$ in a range from 0 to 100 V with a frequency of 50 kHz.
  }
  \label{fig4}
\end{figure}

To experimentally characterize the SHG properties of the metasurface, we used a tunable Ti:sapphire femtosecond laser (Maitai, Spectra-Physics, 80 MHz, 230 fs) to excite the sample, as shown in Figure 4a. The FW light power was adjusted by a combination of a half wave plate (HWP) and a Glan-Taylor prism (GTP). The pump light, polarized along the $z$-axis, was focused onto the sample through a $10\times$ objective lens (NA = 0.25). The SHG light was collected by another objective lens ($100\times$, NA = 0.90) in the transmission direction. A band pass filter (BG40 color glass) was employed to block the pump light, ensuring that only the SHG light reached a balanced detector (PDB220A2/M, Thorlabs). An optical chopper was positioned in front of the detector, and its signal was recorded using a lock-in amplifier (Zurich Instruments MFLI). To elucidate the relationship between the SHG magnitude and the transmission resonance, we measured the SHG strength by sweeping the pump wavelength between 880 to 930~nm with a step of 1~nm. The average pump power was maintained at 30~mW during pump wavelength adjustments, corresponding to a peak intensity of approximately 2.1~GW/cm$^{2}$. As shown in Figure 4b, the SH strength (recorded as the detector output voltage $U_2$) exhibited a distinct wavelength dependence, featuring a pronounced peak around 451~nm, coinciding with the fundamental frequency transmission resonance of 902~nm shown in Figure 3.

For the electrical modulation of SHG, we employed an arbitrary waveform generator (Agilent 33250A) to produce a square alternating voltage signal at a frequency of 50 kHz. This signal underwent amplification to 100 $V_{\text{pp}}$ (peak-to-peak magnitude, -50 V to +50 V output voltage) using a high-voltage amplifier (New Focus 3211) before being applied to the electrodes. The optical chopper was removed from the setup. The electrically induced SHG changes ($|\Delta U_2|$) are depicted in Figure 4c. Notably, the $|\Delta U_2|$ curve exhibits an M-shape, with peaks observed at 448~nm and 454~nm. We further investigated the modulation depth, defined as $|\Delta U_2|/U_2$. The wavelength dependence of $|\Delta U_2|/U_2$ is shown in Figure 4(d), reaching its maximum value of 11.3\% at 458~nm. In Figure 4(e), the changes in $|\Delta U_2|/U_2$ at wavelengths of 454~nm (blue line) and 458~nm (orange line) are presented as a function of the driving voltage $V_{\text{pp}}$, both exhibiting a linear relationship.

\section{Discussion}
In conclusion, we demonstrated a novel approach to electrically modulate SHG from an LN metasurface, leveraging the EO effect. By fabricating a nanohole array metasurface on a thin LN film and applying a specific electric field, we were able to alter the refractive index of the material, causing a shift in resonances and consequently modulating the SHG signal intensity at a specific wavelength. Our findings provide valuable insights for the development of electrically tunable nonlinear light sources, quantum optics, dynamic nonlinear holography, and nonlinear information processing. This work opens new avenues for the practical implementation of electrically controlled nonlinear optical devices, with implications for various fields including telecommunications, imaging, and sensing.  \\

\medskip
\noindent\textbf{Data availability}\par
All relevant data supporting the results of this study are available within the article and its supplementary information files. Further data are available from the corresponding authors upon request.

\medskip
\noindent \textbf{Acknowledgements} \par
This work was supported by National Natural Science Foundation of China (12222408, 92050114, 12174202, 12074200, 12304423, 12304424); National Key R\&D Program of China (2023YFA1407200, 2022YFA1404800, 2019YFA0705000); Guangdong Major Project of Basic and Applied Basic Research (2020B0301030009); China Postdoctoral Science Foundation (2022M721719, 2022M711710); 111 Project (B23045); PCSIRT (IRT0149); Fundamental Research Funds for the Central Universities. We thank Nanofabrication Platform of Nankai University for fabricating samples.

\medskip
\noindent \textbf{Author Contributions} \par
Z.H. and L.Q. contributed equally to this work. L.Q. and M.R. conceived and performed the design. Z.H., L.Q., W.W., J.L., J.Y., C.J., L.B. and Z.G. fabricated samples. Z.H., L.Q., W.L. and C.W. carried out numerical simulations performed optical measurements. Z.H., L.Q., M.R. and W.C. analyzed data. M.R. and J.X. organized and led the project. All authors discussed the results and prepared the manuscript.

\medskip
\noindent \textbf{Competing interests} \par
The authors declare no conflict of interest.

\bibliography{Ref.bib}

\end{document}